\documentclass[reprint, amssymb, amsmath, onecolumn, notitlepage]{revtex4-1}

\usepackage{graphicx}
\usepackage{dcolumn}
\usepackage{bm}

\begin{document}

\title{Deep Hybrid Scattering Image Learning}%

\author{Mu Yang}
 \thanks{These two authors contributed equally to this work.}
\author{Zheng-Hao Liu}
 \thanks{These two authors contributed equally to this work.}
\author{Ze-Di Cheng}
\author{Jin-Shi Xu}
 \email{jsxu@ustc.edu.cn}
\author{Chuan-Feng Li}
 \email{cfli@ustc.edu.cn}
\author{Guang-Can Guo}
\affiliation{CAS Key Laboratory of Quantum Information, University of Science and Technology of China, Hefei 230026, People's Republic of China}
\affiliation{CAS Center For Excellence in Quantum Information and Quantum Physics, University of Science and Technology of China, Hefei 230026, People's Republic of China}

\date{\today}%

\begin{abstract}
A well-trained deep neural network is shown to gain capability of simultaneously restoring two kinds of images, which are completely destroyed by two distinct scattering medias respectively.
The network, based on the U-net architecture, can be trained by blended dataset of speckles-reference images pairs. 
We experimentally demonstrate the power of the network in reconstructing images which are strongly diffused by glass diffuser or multi-mode fiber.
The learning model further shows good generalization ability to reconstruct images that are distinguished from the training dataset. 
Our work facilitates the study of optical transmission and expands machine learning's application in optics.
\end{abstract}

\maketitle


\section{Introduction}

Scattering medium such as biological tissue and seawater would lead to deformation of light beams, which poses an serious limitation in many optical applications. 
The far-field image will be diffused into a speckle pattern after a strong scattering medium. 
In the past decade, various methods has been proposed, such as wavefront shaping~\cite{He,Katz}, optical phase conjugation~\cite{Yaqoob,Si} and transmission matrix (TM) measurement~\cite{Popoff}. However, all these traditional methods meet some shortcomings. Wavefront shaping is based on optical memory effect, which limits the field of view. Phase conjugation and TM measurement with complicated imaging devices have to carry out complex calibration and scanning processes. 


The development of machine learning attracts great attention and provides new ideas for solving the problem of scattering image reconstruction. $Horisaki$ et al.~\cite{Horisaki} used the support vector regression (SVR) to recover face images through scattering media by pixels sampled from the speckle pattern. Artificial neural network has also been used to reproduce the complex amplitude of the scattered light~\cite{Kamilov}. These tasks require experimental data pretreatment, such as sampling, rather than an end-to-end, image to image learning process. 
Recently, deep neural network (DNN), one of the deep architectures of broader family of machine learning methods, has been used in the investigation of optical imaging, such as lensless computational imaging~\cite{Sinha}, ghost imaging~\cite{Shimobaba}, wavefront sensing~\cite{Paine} and turbulence correction~\cite{Liu}. 
Combining with several special treatment, the famous network structures, such as convolution deep neural network (CNN) and recurrent deep neural network (RNN) have been created and widely used in image classification~\cite{Chan,Ciresan}, face recognition~\cite{Parkhi} and pattern detection~\cite{LeCun Y,Szegedy}. DNN training can be thought as an approximation of generic function which can character a physical process. The training process depends on a large number of samples often called "big data", aims to generate a computational architecture that accurately maps all inputs to their outputs. Image reconstruction and optimization can be transformed into a supervised learning problem, which aims to find the mapping relationship between the original images and the graphics though a ruinous optical system.

In this paper, we use an end-to-end DNN of semantic segmentation, to achieve images reconstruction, in which they are diffused by strong scattering medium and a multi-mode fiber (MMF). Some work using similar settings were carried out recently~\cite{S0,S1,S2,S3}.
Comparing to these works, a major advantage for our work is that the neural network has ability to simultaneously reconstruct two distinct kinds of speckle images, without further need of independent training.
The structure of DNN is U-net~\cite{Ronneberger15} architecture which
consists of two parts: a contracting path to obtain context information and a symmetric expansing path for precise positioning. There are three advantages of this network: (i) It supports learning from only small amount of data; (ii) By classifying each pixel, a higher segmentation accuracy is achieved; (iii) It is fast to divide the images with the trained model.
Our training set are high resolution handwritten numbers based on the famous data set $MINST$. The network is shown to well reconstruct handwritten numbers. The trained network has strong generalization ability, which can be further used to reconstruct handwritten letters. Moreover, it can also be used to reconstruct speckle images passing though different positions of a diffuser, which means what learned by DNN is a hidden scattering model, rather than only a mapping function.

\section{Learning Scheme}

As for optical progressing though scattering medium and noisy channels, the physical process can be described as $\hat{I}(x,y)=\mathcal{F}(I(x,y))$, where $I(x,y)$ and $\hat{I}(x,y)$ represent original images and scattered images respectively. The general aim of image reconstruction is to find the inverse function $\mathcal{F}^{-1}$. However, the function always has a large number of undetermined parameters and implicit variables, so that it's unable to find the functional expression directly. Therefore, we usually look for an approximate description $\mathcal{H}\to\mathcal{F}^{-1}$ to obtain an result similar to the original image $I^{'}(x,y)=\mathcal{H}(\hat{I}(x,y))\approx I(x,y)$. The closer $I^{'}(x,y)$ to $I(x,y)$, the better effect will be. Fortunately, DNN can construct arbitrary nonlinear mapping $\mathcal{H}_{\theta}$ by statistical means. The function parameters $\theta$ are learned from large amount of samples.

In our work, we adopt a scheme including learning the mapping function $\mathcal{H}_{\theta}$ from "speckle-object" pairs and predicting speckle patterns. 
The computational imaging system consists of three parts, which are data preprocessing section, optical system, and the neural network. In the training progress,  we first load the original image from the data set $MINST$. Since the resolution of the original dataset is $28\times28$ pixels, we need to expand the resolution to $512\times512$ pixels by $upsampling$, so that it can be coupled into the optical system. Then, these images are sent to different optical channels to get speckle images and object images respectively. A large number of "speckle-object" pairs are used to train DNN. In the test progress, we load test set (not the same as training set) in the same way, and only get the speckle images from optical scattering channels. The well trained DNN is then used to reconstruct the non-interfering images with these speckle patterns.

\begin{figure}[htbp]
	\centering
	\includegraphics[width=1\linewidth]{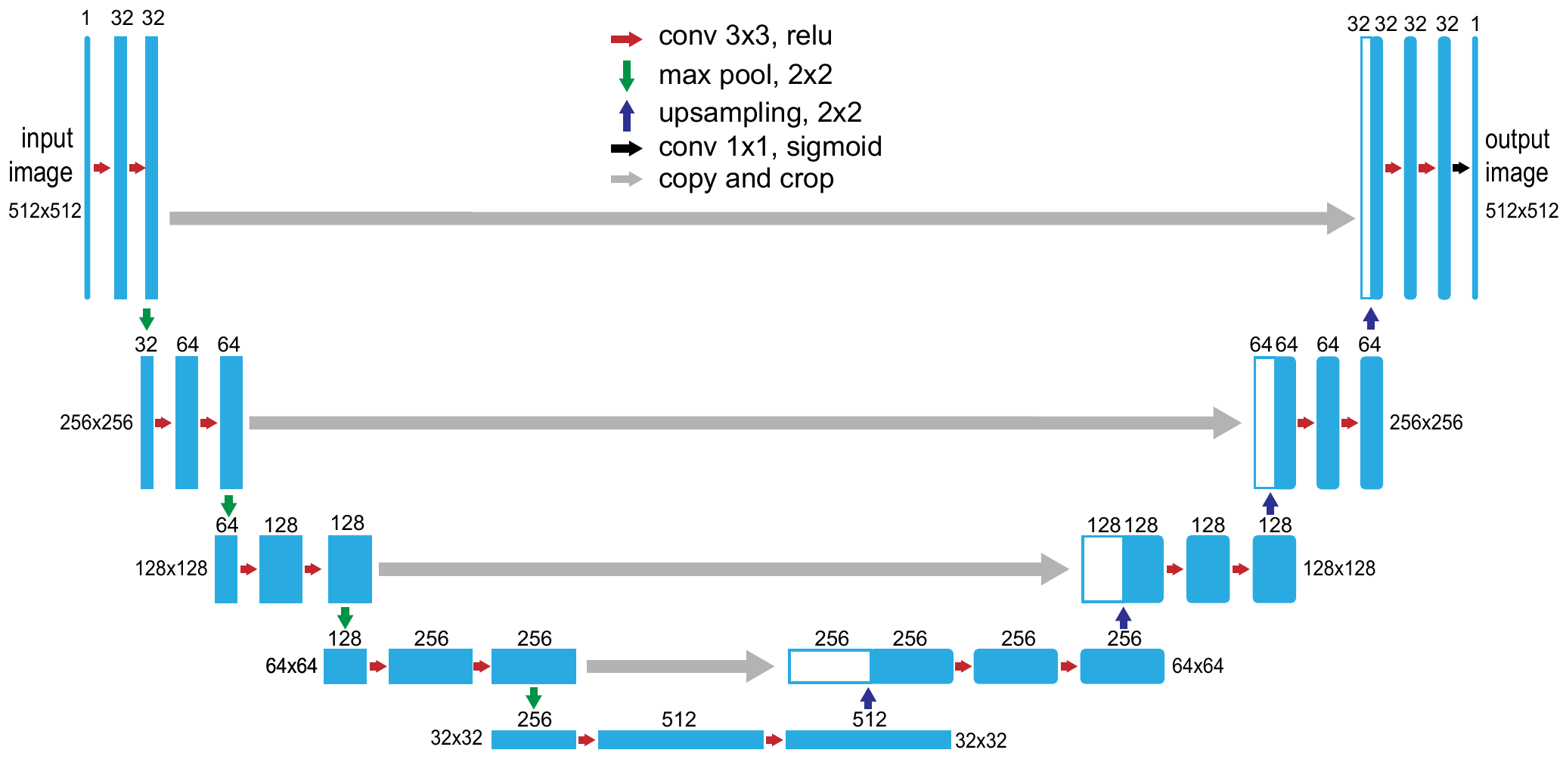}
	\caption{The U-net architecture of DNN
	}
	\label{U-net}
\end{figure}

The U-net architecture of DNN is shown in Fig.~\ref{U-net}. The input layer and output layers is of $512\times512$ pixels which is corresponding to the pictures from the optical system. The U-net consists of two paths. In the contracting path, there are five blocks and each block comprises two convolution layers followed by a maximum pool layer. In the expanding path, it connects five blocks and two convolution layers are followed by an upsampling layer. Some of the convolution layers in the contracting path merge with the corresponding convolution layers in the expanding path to avoid the loss of pixels. The input image is eventually transformed into another image though this DNN, so it's an end-to-end progress. Following conventions of utilizing U-net architecture, the optimizer is Adam~\cite{kingma14} and the loss function is the binary cross entropy, which is defined as $-\dfrac{1}{n\times N}\sum_{k,i,j}[y_{k}(i,j)\log a_{k}(i,j)+(1-y_{k}(i,j))\log (1-a_{k}(i,j))]$. $k$ is the sample sequence number, $i,j$ represent pixel sequence number; $y$ represents the intensity of reference images; $a$ is the output images of DNN; $n$ is the number of all pixels of each image and $N$ is the number of all samples. 

\section{EXPERIMENTAL SETUP AND RESULTS}

\begin{figure}[htbp]
    \centering
    \includegraphics[width=0.97\linewidth]{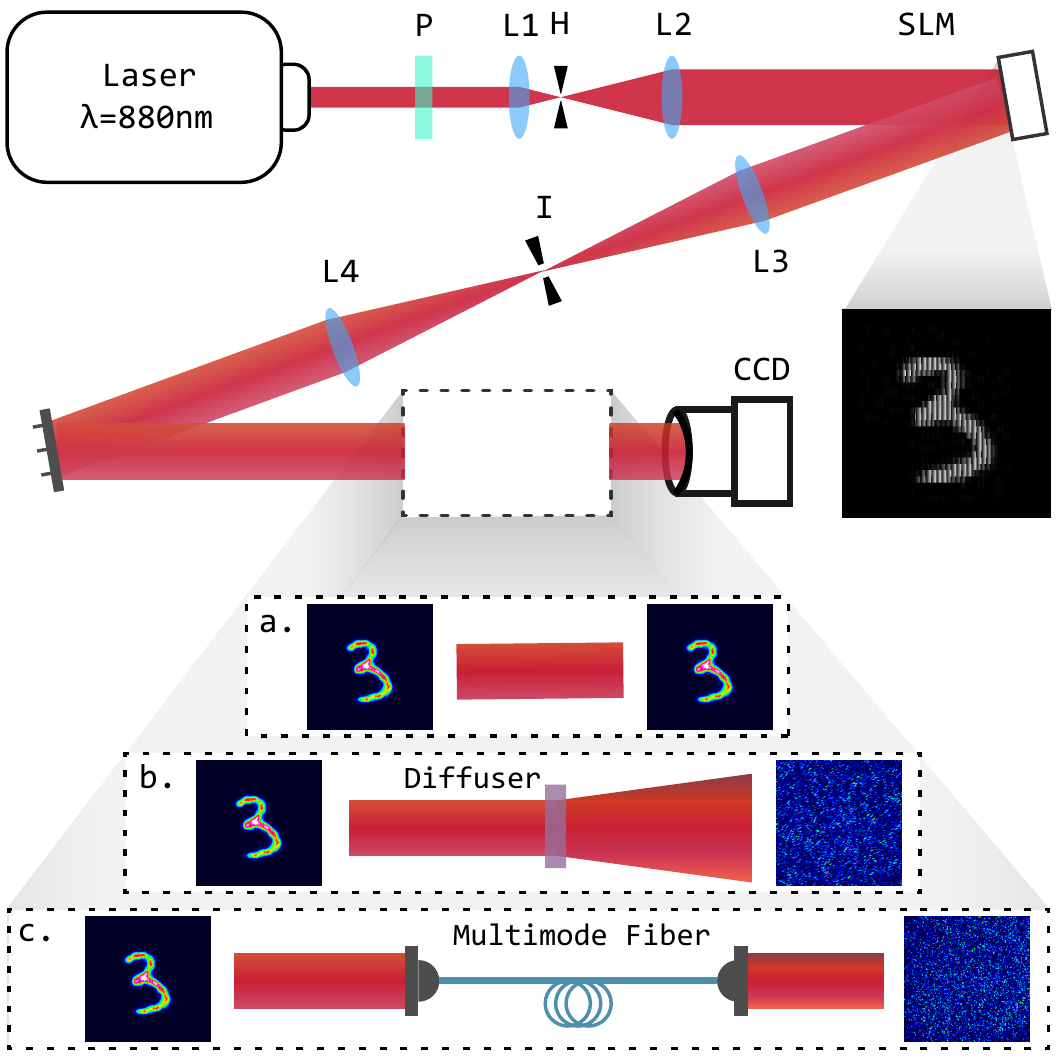}
    \caption{Experimental setup for recording speckle and reference patterns. P: polarizer, L: lens, H: pinhole, I: iris (used for extracting first order diffraction.)}
    \label{fig:exp-sch}
\end{figure}

We generate dataset in the experimental setup, which is shown in Fig.~\ref{fig:exp-sch}.
By virtue of a method demonstrated in prior works~\cite{davis99, leach05, eliot13}, a phase-only spatial light modulator(SLM, Hamamatsu, pixel pitch $20\mu m$) was used to generate desired intensity pattern. 
2000 images of $28\times28$ pixels resolution handwriting numbers were acquired from online database \textit{MINST}, which are then upsampled and rasterized into phase-only computer-generated holograms and displayed on the central $560\times560$ pixels of SLM (the handwritten number 3 is shown as an example).
An expanded $880$nm laser reflected off the SLM and the hologram effectively diffracted the incident beam. The first order diffraction carried an intensity pattern representing the original input image, and propagated through an aperture, while other orders were spatial-filtered by the aperture. 

\begin{figure*}[t]
    \centering
    \includegraphics[width = 0.4\linewidth]{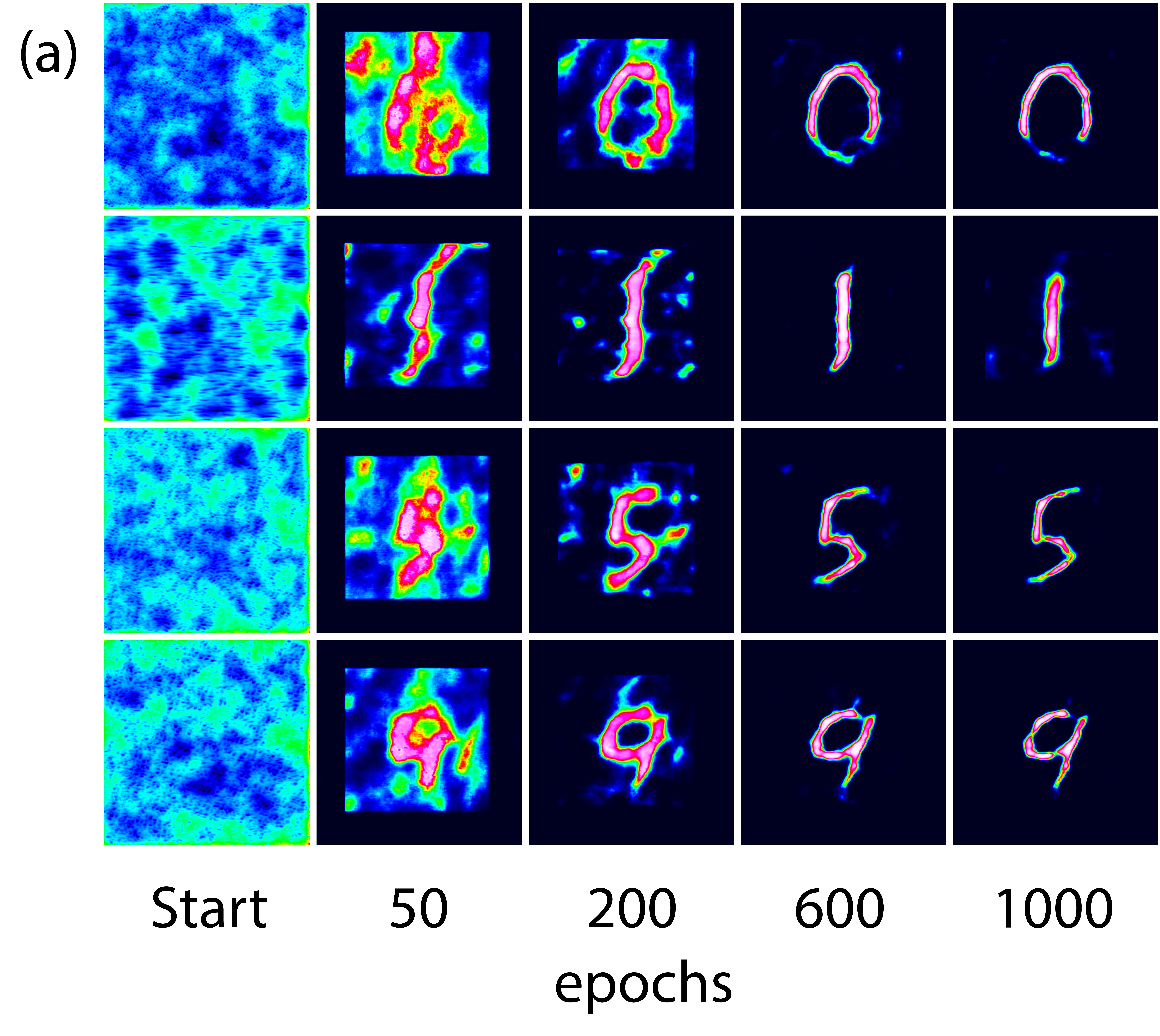}
    \hspace{48pt}
    \includegraphics[width = 0.4\linewidth]{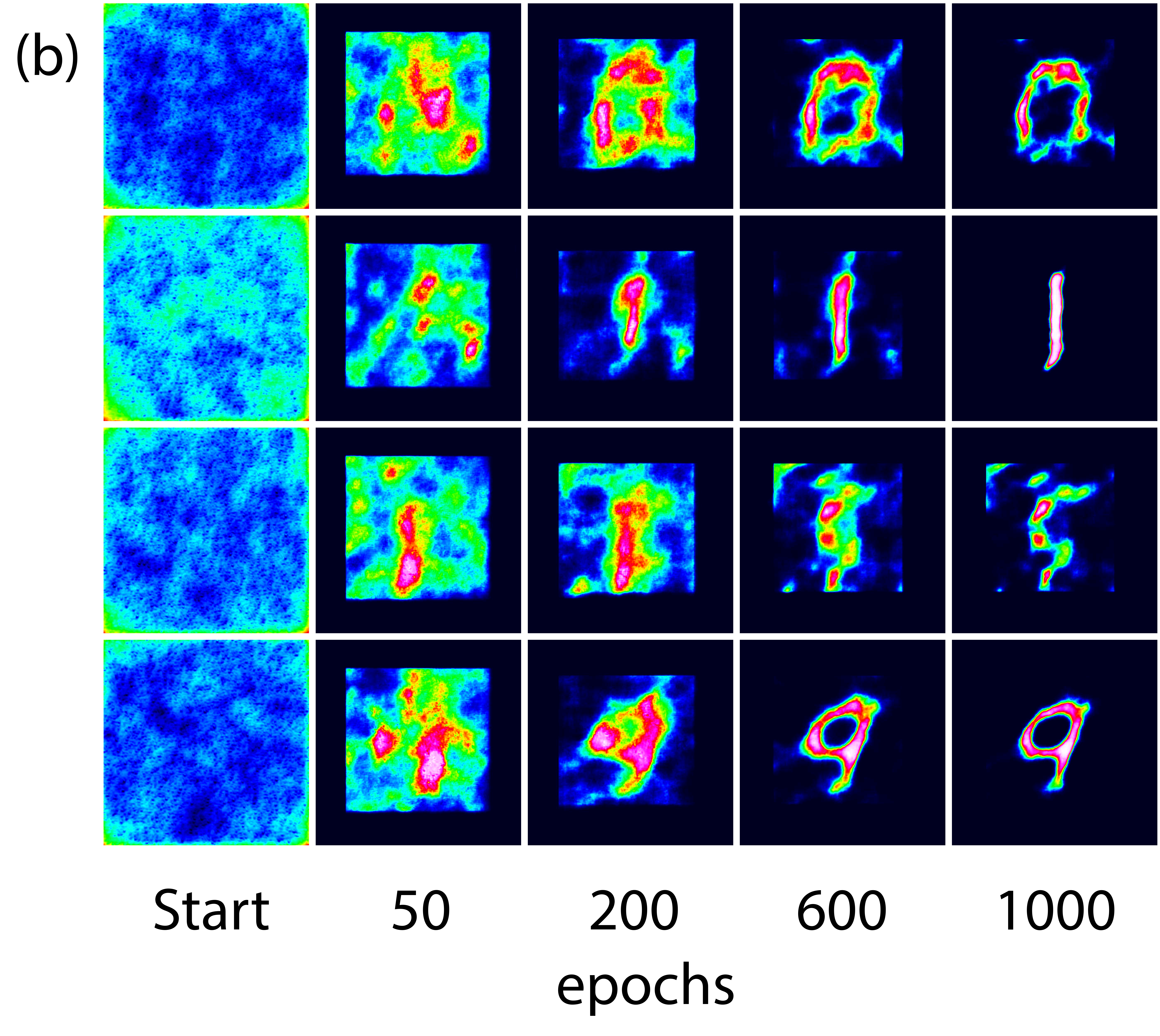}
    \hspace{16pt}
    \caption{Graphical presentating the enhancement of image reconstruction ablilty of the neural network trained by blended dataset. (a) and (b) correspond to evolution of reconstructed speckle patterns created by glass diffuser and MMF respectively, plotted against training epochs.}
    \label{fig:exp-rst}
\end{figure*}

\begin{figure}[htbp]
    \centering
    \includegraphics[width = 0.97\linewidth]{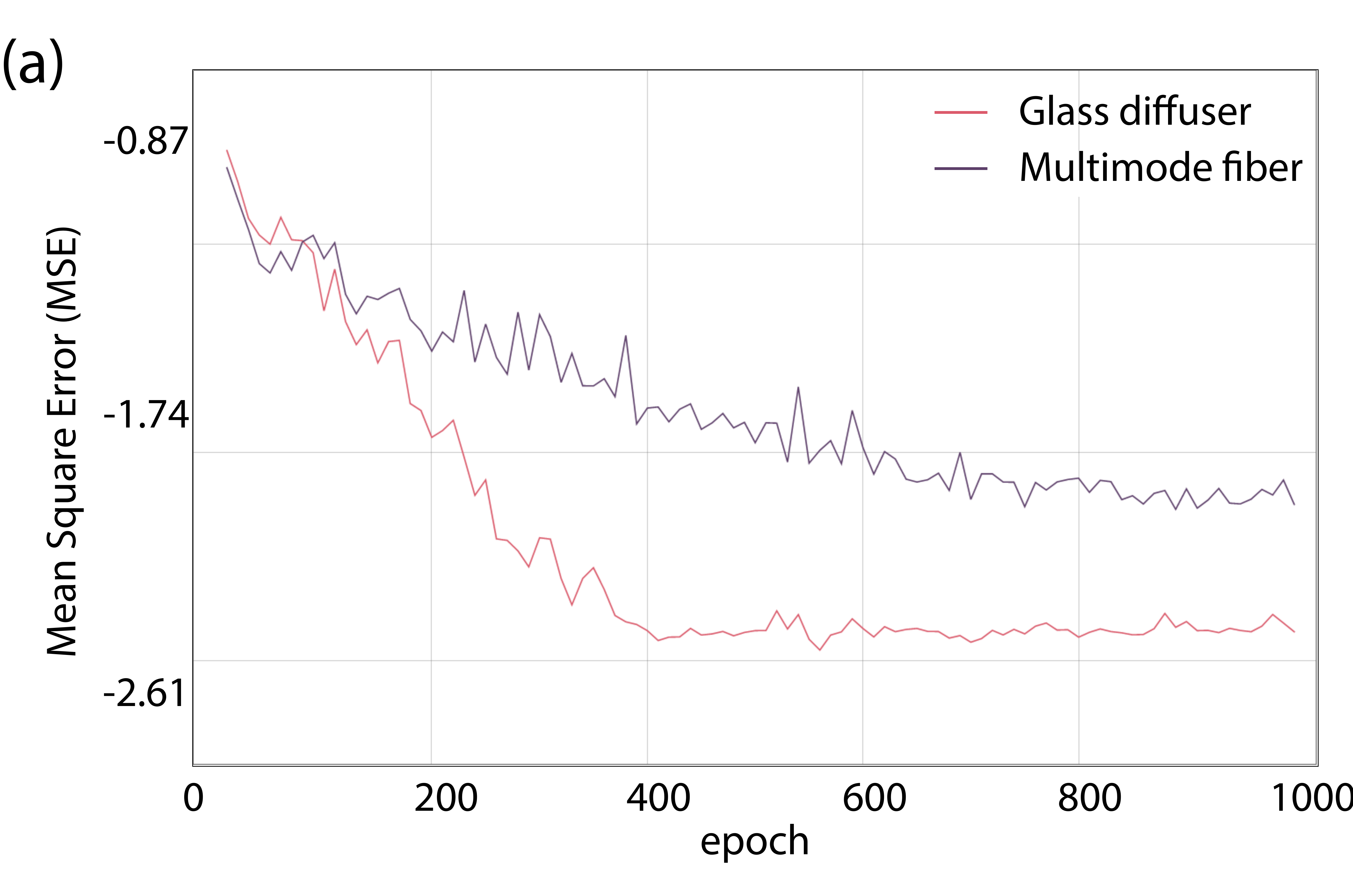}
    \includegraphics[width = 0.97\linewidth]{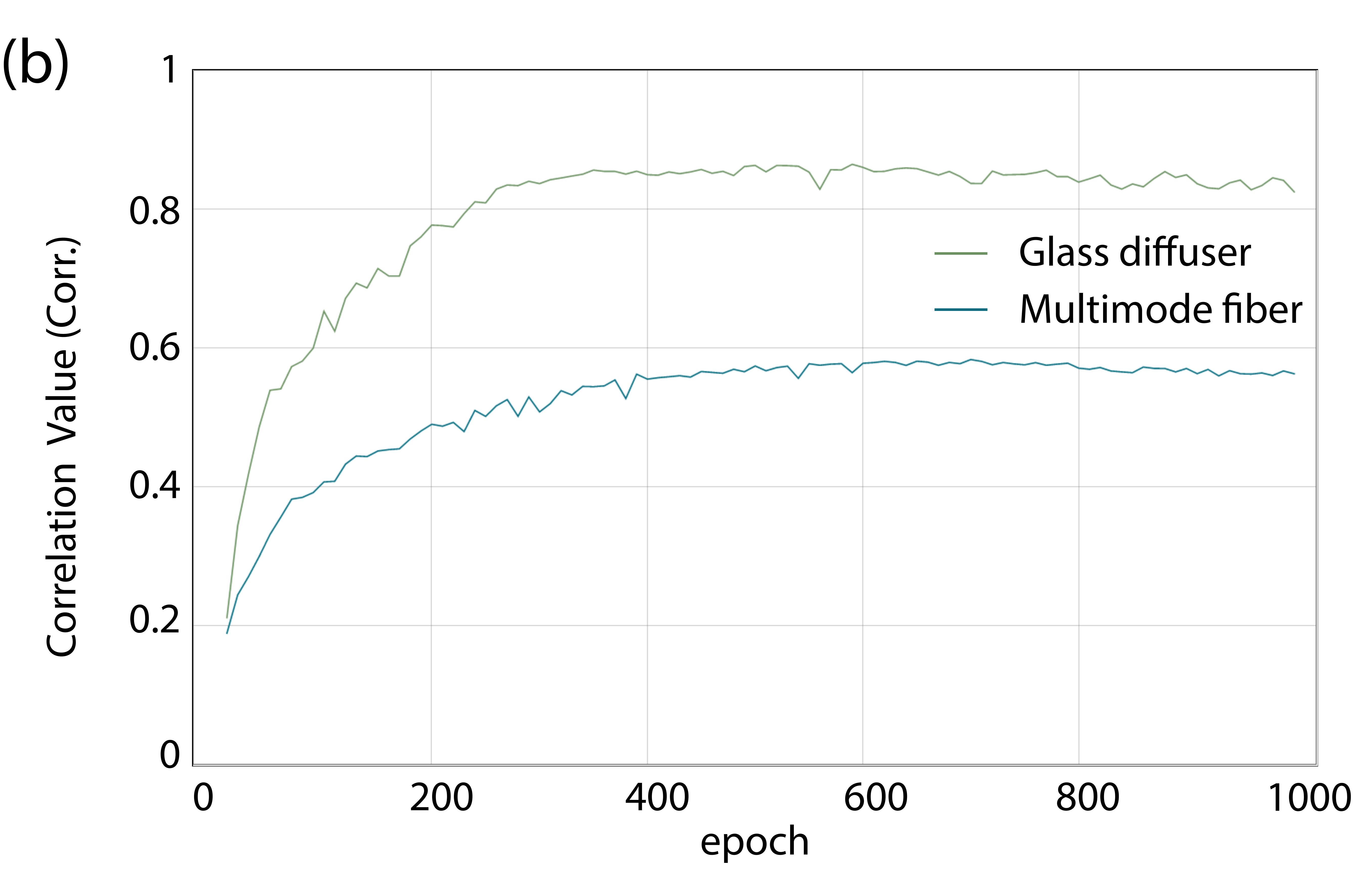}
    \caption{Quantified performance of the neural network with respect to epoch. (a) Logarithm of mean square error (MSE). (b) Correlation of reconstructed pattern and the original image. Plot legends correspond to speckle pattern types.}
    \label{fig:cal}
\end{figure}

Three imaging channels are used: (a) free-space transmission; (b) a glass diffuser plate (Thorlabs, 220 Grits), installed on a motorized rotation mount for future usage, whose angle was fixed during the whole training process, and aligned through the principal optical axis; and (c) a multi-mode optical fiber (Ideaoptics, $L=1$m, core diameter $600\mu$m, $N.A.=0.22$), respectively. Patterns collected in setup (a) is used as ground truth labels, and output beams in setups (b) and (c) forms different kinds of speckle patterns. 
Each images collected at the CCD camera were normalized and saved in $512\times512$ pixels size to match the scale of the neural network.
The SLM iterated over 2000 kinoforms to create corresponding images.

In training progress, ``object-spekle'' pairs from (a)/(b) and (a)/(c) were mixed up to form a hybrid training data set and then were sent to DNN modle.
The training was conducted on a computer with two NVIDIA GTX 1080Ti graphics processors. The U-net was built on TensorFlow API framework. Using the Adam optimizer with a characteristic learning rate to be $10^{-6}$, treatment model was iteratively fitted for 1000 epochs and control model was fitted up to 100 epochs.
The training process elapsed 2 days.
After each fitting epoch, predicting model was saved and ten test speckle images (different from the training set) from each kind of scattering were sent into the U-net to make prediction. Fig.~\ref{fig:exp-rst} shows some of the forecasting results presented by DNN at the beginning, 50, 200, 600 epochs and the end of training. 
The graphical results suggested that the neural network was capable of reconstructing original pattern from both kind of scattering images.
 
To quantitatively analyze our hybrid model, we first introduce mean squared error (MSE) and correlation value (Corr.) A grayscale picture $A$ can be represented by matrix $\bm{A}$, whose element $A(i, j)$ represents luminance at pixel $(i, j)$, and average luminance of whole picture is represented by $\bar{A}$. Given total pixel numbers N, MSE and Corr. of two pictures are respectively defined as`
\begin{subequations}
\begin{align}
    MSE(\bm{A}, \bm{B}) &= \frac{1}{N} \sum_{i, j} (A(i, j) - B(i, j))^2, \\
    Corr(\bm{A}, \bm{B}) &= \nonumber\\
    & \frac{\sum_{i, j} (A(i, j) - \bar{A})(B(i, j) - \bar{B})}{\sqrt{\sum_{i, j} (A(i, j) - \bar{A})^2\times\sum_{i, j} (B(i, j) - \bar{B})^2}}.
\end{align}
\end{subequations}

Comparison between original and reconstructed patterns were made by calculating their MSE and Corr. values. 
For two patterns which closely represent each other, MSE should vanish and Corr. should approach unity. 
Fig.~\ref{fig:cal} shows tendency of MSE and Corr. of reconstructed (by treatment model) and reference patterns, averaged over ten testing samples, plotted against training epochs. 

During an 1000 epochs training, the reconstructed patterns showed progressively closer resemblance to the reference patterns. 
At the end of training process, MSE between reconstructed and reference patterns, corresponding to diffuser and MMF scattering medias, fell below $3\times10^{-3}$ and $10^{-2}$, respectively.  Moreover, correlation value between recovered and reference patterns in glass diffuser and MMF experiments capped at about $0.87$ and $0.58$ respectively.
On the contrary, the DNN model trained by only glass diffuser patterns fails to reconstruct MMF diffused patterns, and vice versa. 
In conclusion, we only use one neural network to achieve two types of images restoration simultaneously. 

\begin{figure}[htbp]
    \centering
    \fbox{\includegraphics[width = 0.97\linewidth]{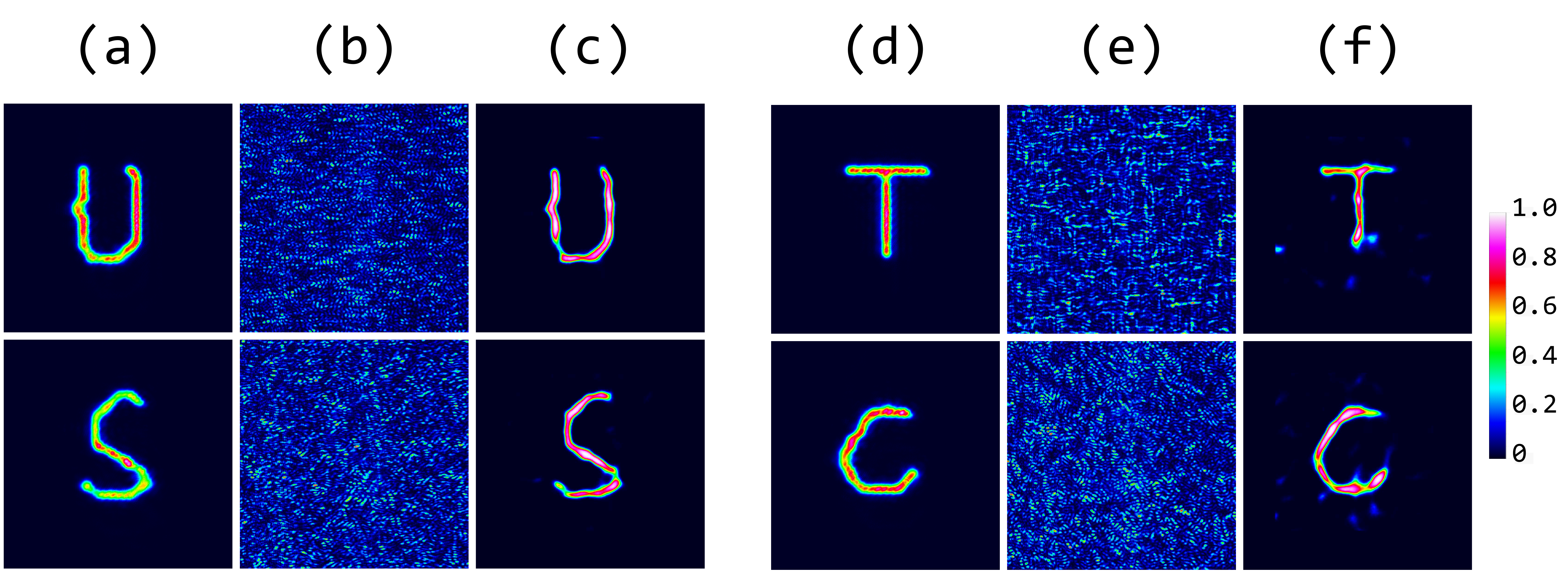}}
    \caption{Reconstructing pattern of letter through the network trained by numbers. (a) and (d) Original patterns without distortion. (b) and (e) The corresponding speckle patterns measured after propagating through a frosted glass. (c) and (f) U-net reconstructed patterns from (b) and (e), respectively.}
    \label{fig:exp-let}
\end{figure}


We further demonstrated that, although the training dataset, MINST, is made of handwriting numbers, after being trained by sufficient volume of data, the DNN also gained capability to reconstructing patterns of simple symbols. 
As shown in Fig.~\ref{fig:exp-let}, DNN forecasted handwritten letters from speckle images of frosted glass, with MSE and Corr. being $2.65\times10^{-3}$ and $0.89$ respectively, which suggests that these two data sets are unlikely, in the machine view, strictly orthogonal or irrelevant. 

\begin{figure}[htbp]
    \centering
    \fbox{\includegraphics[width = 0.97\linewidth]{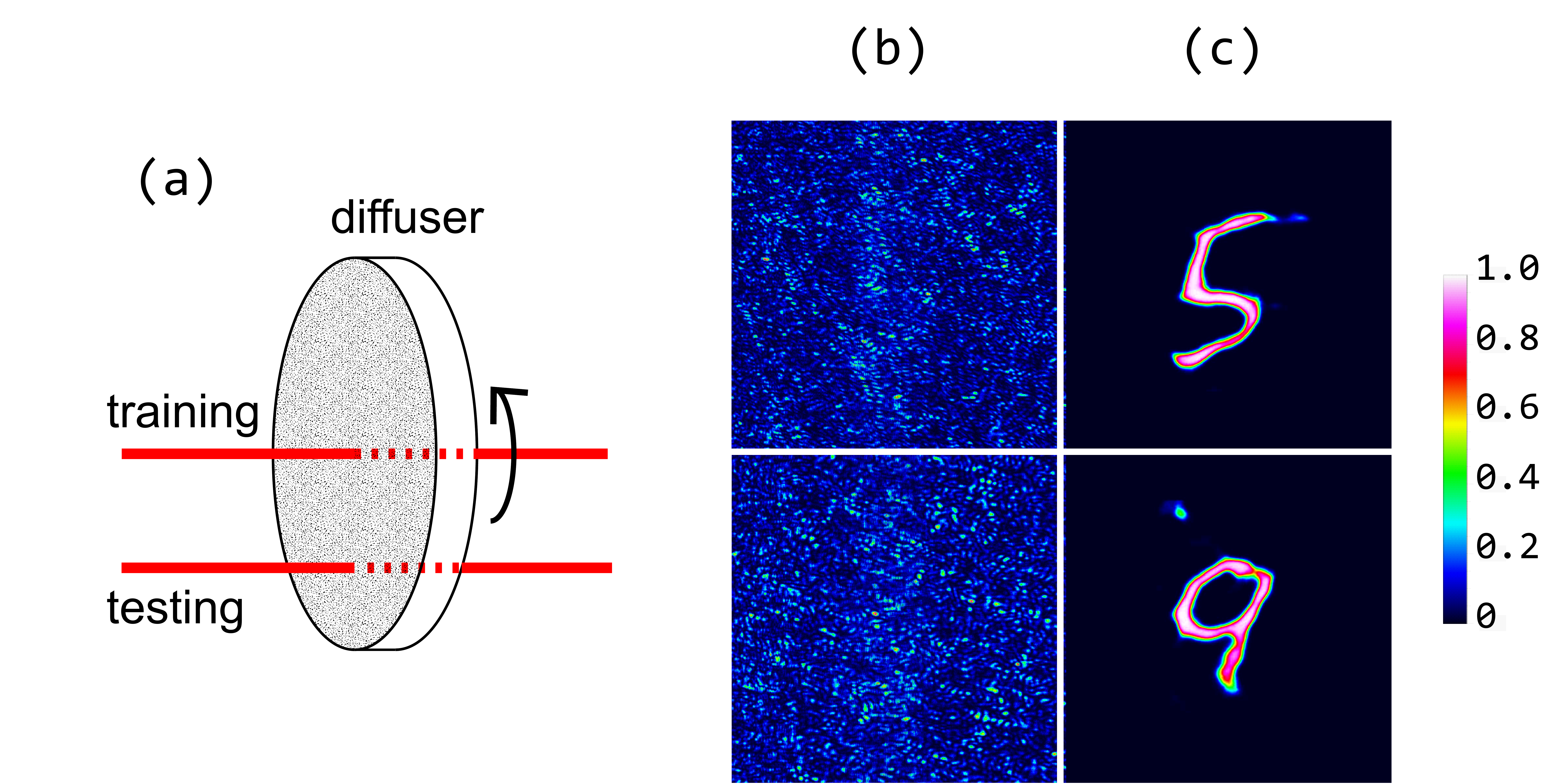}}
    \caption{Reconstructing patterns using speckle images from different location on the diffusing plate. (a) The setup of generating predicting data set. Each of the test image passes through different off-axis area of the frosted glass mounted on a motorized rotation stage. While the training image passes through the optical axis which is same as that shown in fig.~\ref{fig:exp-rst}. (b) The corresponding speckle images of 5 and 9. (c) U-net reconstructed patterns from (b).}
    \label{fig:exp-ran}
\end{figure}

Another trial of model generalization ability was made by shifting position of scattering plate in the test process. During acquisition of test data set, the diffuser was rotated $13^\circ$ between each tested image, rendering the scattering process somewhat different across each measurement. 
The model we previously built, in spite of absence of detailed information about peripheral parts of diffuser during training, still exhibited the ability of predicting speckle images generated though random parts of the same diffuser, see Fig.~\ref{fig:exp-ran} with two test numbers, 5 and 9, for example. 
This cross-data-set reconstruction attained MSE and Corr. of $7.0\times10^{-3}$ and $0.78$ respectively.
This result suggests that DNN not simply learns the mapping matrix of specified location in the diffuser channel, but learns the scattering mechanism of the diffuser itself.

\section{CONCLUSION}

Focusing more deeply on generalization ability of the network, we successfully achieve reconstructing two kinds of scattering images by only one neural network, and demonstrated the generalization ability of neural network to solve optical transmission problems.   
Using a blended trainig dataset, The MSE for reconstructed images though the scattering glass and multi-mode fiber are made less than $10^{-2}$. The correlations of reconstructed images and original images are over $0.87$ and $0.58$ respectively. We also obtained several interesting conclusions, that the trained model can recover some untrained kinds of patterns, e.g., reconstructing images of handwritten Latin letters, and patterns scattered by diffusers of the similar attributes. 

DNN for the end-to-end image restoration problem has regression converging property and great generalization ability. Our results suggest that the solution of more complex problems can be obtained with the help of deep learning. 

This work was supported by the National Key Research and Development Program of China (Grant No. 2016YFA0302700), the National Natural Science Foundation of China (Grants No. 61725504, 61327901, 61490711, 11774335 and 11821404), the Key Research Program of Frontier Sciences, Chinese Academy of Sciences (CAS) (Grant No. QYZDY-SSW-SLH003), Anhui Initiative in Quantum Information Technologies (AHY060300 and AHY020100), the Fundamental Research Funds for the Central Universities (Grant No. WK2470000020 and WK2470000026).



\end{document}